# Doppler Effect of Time and Space

Giovanni Zanella

*Dipartimento di Fisica e di Astronomia dell'Università di Padova and Istituto Nazionale di Fisica Nucleare, Sezione di Padova, via Marzolo 8, 35131 Padova, Italy*

**Abstract**

*This paper shows as the relativistic Doppler effect can be extended also to time and space associated to moving bodies. This extension derives from the analysis of the wave-fronts of the light emitted by a moving source in inertial motion in the empty space, as viewed from the stationary reference. Indeed, time and space can be represented by the same vector quantities, which appear asymmetrical in forward and back direction along the path of the moving body. Consequently, the whole size of the moving bodies dilates along the direction of their motion, as their path. Thought experiments and real facts demonstrate this issue.*

## 1. Introduction

A. Einstein, in his special theory of relativity (SR) [1] considered the relativistic motion of a source of "electrodinamic waves" by showing the change of frequency and of energy for the waves perceived by a stationary observer (*Relativistic Doppler Effect*), but he didn't ascribed this effect also to time and space.

Einstein's view of the time was based on the assumption of mechanical clocks positioned in any point of the space and synchronized by rays of light, tacitly assuming the path of the rays of light as the universal time reference. However, the use of mechanical clocks as time measurers is not physically suitable, because they furnish a scalar measurement which does not evidence the connection of the time with the space, as it appears in light clocks.

The contraction of the sizes of the moving bodies along the direction of their motion was postulated by G.F. FitzGerald and H.A. Lorentz to explain the negative output of the Michelson-Morley experiment and to rescue the hypothesis of a stationary ether [2][3].

A. Einstein, in his special theory of relativity (SR) [1], derived the *Lorentz transforms* (LT) completely removing the hypothesis of the ether and, applying LT, he demonstrated both size contraction in moving bodies and time slowing in moving clocks. But, if the time slowing in moving clocks is experimentally proved, the contraction of the bodies has been always object of discussion and it is not been tested with real experiments and of thought. See, for example, the papers of J. Franklin [4], S.D. Agashe [5], R.D. Klauber [6] or of Y. Pierseaux [7] and their references.

In this paper is argued, starting from the principles of SR, that the bodies in motion undergo a dilation along the direction of their motion together to their path, as well the time associated to their.

We recall here the two known principles of the SR [1]:

1. *The laws of the physics are the same no matter if referred to one or the other of two systems of co-ordinates in uniform translatory motion.*
2. *Any ray of light moves in the stationary system of co-ordinates with determined velocity (c) whether the ray is emitted by a stationary or by a moving body.*

Hence, $\quad c = $ *light path / time interval*,

so $\quad$ *time interval = light path / c*

and putting conventionally $c=1$, the *light path* expresses the *time interval*.

It is worth to underline that the equivalence *light path = time interval* furnishes in turn the constancy of *c*, provided that the time interval be identified on the same path of the ray of light and measured from a same reference system, otherwise the constant *c* cannot be assured. We define this time as *light-time* and use for it the symbol $t_l$.

We shall found our analysis also on these further hypotheses, implied in SR:

1. *The empty space is isotropic and homogeneous.*
2. *The empty space is not a "stationary" support of physical events.*
3. *The physical events occurring in the same spatial point and in the same instant (coincident events) are inseparable, in time and space, if viewed from any other reference system.*
4. *The closed wave-front of propagation of a radiation, in the empty space, or the closed outline of a rigid body, are closed in respect to any other reference system.*

## 2. Light-time or clock-time?

Analyzing the operation of the transversal light clock (Fig.1), in motion with the inertial velocity *u* (clock 2) in comparison to an equal stationary clock (clock 1), we see the necessity to define two types of times.

Looking to the triangle ABC of clock 2 (Fig.1) the *light-time* between two consecutive reflections is amplified by a factor *γ*, in comparison to the corresponding time *t* of clock 1, so we have

$$(u\gamma t)^2 + c^2 t^2 = c^2 (\gamma t)^2 \quad , \qquad (1)$$

where $\gamma = 1/\sqrt{1 - u^2/c^2}$ is the *Lorentz factor*.

Now, if in clock 1 the path $t_l$ of the light ray (between two or more reflections) is directly proportional to the corresponding *number of reflections* $t_c$, in moving clock 2 the relation between $t_l$ and $t_c$ is inverse and depends on the velocity *u* by the *Lorentz factor*. Indeed, increasing the velocity *u* of clock 2, the *light-time* increases, while $t_c$ decreases relatively to $t_l$. We define the time $t_c$ as *clock-time*.

Obviously, in the path AB the ratio of the dilated path *cγt* with the dilated time *γt* furnishes the velocity *c*, so also the velocity *u* corresponds to the ratio of the dilated path *uγt* with the dilated time *γt*.

On the contrary, the ratio of these dilated paths with the time $t_c$ would furnish the velocity *γc* and the velocity *γu*. But, being experimentally proved that the rate of the ticks of any moving clock slows down relatively to an identical stationary clock, and also for the principle of the constancy of the velocity of the light, the universal time reference is necessarily the *light-time*. Hence, the velocities *γc* and *γu* do not exist in the reality.

In general, we can think that the *light-time* associated to any moving body dilates by the *Lorentz factor*, like its path.

## 3. Doppler effect on time and space

Supposing isotropy and homogeneity of the empty space, as well as the impossibility to assume the empty space as a stationary reference, Fig.2 shows spherical and concentric wave-fronts emitted by a point light source, as viewed locally. In Fig.2 the solid arrows measure the elapsed time from a starting instant, indifferently on the selected direction. Fig.3 shows instead the wave-fronts emitted from a light source put in the origin of a Cartesian system S', moving with the relativistic velocity *u* according its x'-axis, compared to a stationary system S, which axes are parallel to those of S'. Fig.3 includes the supposed time amplification due to the *Lorentz factor* with the consequent appearance of two times, corresponding to the *forward* and *back* path of the light rays emitted along to the direction of the motion of the source and started when the origins of S and S' were coincident. Hence, also the space covered by the ray of the light assumes two states, one contracted (forward) and the other dilated (back), according to the path of the moving light source.

These two oriented states of time-space can be considered as vectors, whose average length can be deduced by $x_f$ and $x_b$ (or $t_f$ and $t_b$) covered, forward and back, by the light-rays emitted by the moving source. Therefore, looking to Fig.3 and introducing the *Lorentz factor*, we have

$$t_f = \gamma\left(t - \frac{ux}{c^2}\right) = \gamma t\left(\frac{c-u}{c}\right), \qquad (3)$$

intending $t=x/c$, while the corresponding relationship for $x_f$ is

$$x_f = \gamma(x - ut) = \gamma x\left(\frac{c-u}{c}\right) = x\sqrt{\frac{1-u/c}{1+u/c}} \quad . \qquad (4)$$

We can note that, despite the amplification by $\gamma$, $t_f$ and $x_f$ contracts until zero when *u* tends to *c*, while $t_b$ dilates according to

$$t_b = \gamma\left(t + \frac{ux}{c^2}\right) = \gamma t\left(\frac{c+u}{c}\right), \qquad (5)$$

where the co-ordinates are intended in absolute value.

So we can also write

$$x_b = \gamma(x + ut) = \gamma x\left(\frac{c+u}{c}\right) = x\sqrt{\frac{1+u/c}{1-u/c}} \quad , \qquad (6)$$

intending the co-ordinates in absolute value.

In conclusion, the average length of the light path along the direction of motion of the source, will be

$$\langle t \rangle = \frac{1}{2}\left[\gamma t\left(\frac{c-u}{c}\right) + \gamma t\left(\frac{c+u}{c}\right)\right] = \gamma t \quad . \qquad (7)$$

A similar result is worth for the space. Indeed, the lengths $x_f$ and $x_b$ (Fig.3) are represented by the same paths of the light, assuming $c=1$, so we obtain

$$<x> = \gamma x \qquad . \qquad (8)$$

Eq.(7) and Eq.(8) prove as the average distance travelled by the light dilates at the same manner of the time, by the *Lorentz factor*. Fig.3 shows also the impossibility for the light source, and potentially for any body, to reach and overpass the velocity of the light, because in this occurrence the crossings (or the superposition) of the wave-fronts among them would appear also locally. In the following, we will expose some though experiments and real facts which prove further the physical coherence of the dilation of the lengths associated to the moving systems.

**4. Moving rigid shell**

Supposing isotropy and homogeneity of the empty space, as well as the impossibility for the empty space to be a stationary reference, Fig.3 shows spherical wave-fronts emitted by a point light source, integral in the center of a transparent spherical rigid shell, as viewed locally. Therefore, it is impossible that can exist a reference where these wave-fronts intersect among them, or with the shell, because these intersections would appear as such also locally. Consequently, any body (intended as light source) is prevented to reach the velocity of the light, due the impossibility for hypothetical wave-fronts to intersect among them or to be all superimposed in the same instant and in the same point, because these "coincidents events" would be seen also locally.
Fig.4 shows the impossibility of symmetrical contraction (item 1), of unchanging (item 2), or symmetrical dilation (item 3), of the shell of Fig.3, when it is moving relatively to the stationary reference S. Indeed, the inevitable intersections of the wave-fronts of the light with the shell cannot be viewed locally.
On the other hand, Fig.5 shows the impossibility of contraction of the wave-fronts of the light, in concert with the contraction of the shell of Fig.4, due the impossibility of the wave-fronts of the light to be dragged by the motion of the source.
Fig.6 shows instead as the asymmetrical distortion of the moving shell of Fig.4 avoids any intersection of the wave-fronts with the shell itself, also introducing the relativistic dilation.

## 5. Lorentz transforms and Einstein's results

LT is a system of equations which connects the co-ordinates *t*, *x*, *y*, *z*, characterizing an event in S, with the co-ordinates *t'*, *x'*, *y'*, *z'*, characterizing the same event in S'. In particular, if an event has co-ordinates *t*, *x*, *y*=0, *z*=0, then the co-ordinates *t'*, *x'*, *y'*, *z'*, viewed from S, are

$$t' = \gamma\left(t - \frac{ux}{c^2}\right); \quad x' = \gamma(x - ut); \quad y' = 0; \quad z' = 0. \qquad (9)$$

We can note that, if *t=x/c*, Eq.s (9) become Eq.(3) and Eq. (4). In other words Eq.s (9) can describe the "forward" side of the ray-light emitted from a moving source, while the "back" side of light propagation can be expressed by Eq.s (9) if *x* and *t* co-ordinate is intended negative.

In conclusion, if LT converts the time as *light*-time, it cannot convert the time as *clock-time*, because this time inversely behaves to the *light-time*.

About the Einstein's results, he using LT recognizes a shortening of the sizes of the moving bodies, because he avoids the asymmetry of the equation $x' = \gamma(x - ut)$ putting *t*=0 and *x'* to a steady positive value. Indeed he affirms [1] that: ... *the equation of the surface of a rigid sphere of radius R, at rest relatively to the moving system* S' *and with its centre at the origin of co-ordinates of* S'... *expressed in x,y,z at time t=0, is* $\gamma^2 x^2 + y^2 + z^2 = R^2$ . *A rigid body which, measured in a state of rest, has the form of a sphere, therefore has in a state of motion - viewed from the stationary system - the form of an ellipsoid of revolution with axes R/$\gamma$ , R, R. Hence, ... the X dimension appears shortened in the ratio 1:1/$\gamma$*.

In conclusion, Einstein ignores the Doppler effect on time and space, when he speaks of a *shortened ellipsoid of revolution*.

About the time, Einstein operates with LT on the "forward" side and he interprets the shortening of the *light-time* as slowing of a *clock-time*. Indeed, he affirms [1] ... *that the time marked by a clock, at rest relatively to the moving system, when viewed from the stationary system is slow* by 1-1/$\gamma$ *seconds per second*. So Einstein unintentionally goes in accord with the comprehensive dilation of the *light-time*, which involves the slowing of the *clock-time*.

## 6. Moving longitudinal light clock

The necessity of the dilation of the sizes of moving bodies appears in a acute form in the longitudinal light clock, sketched in Fig.8.

Supposing a light source put in the origin of the previous system S', the light propagation along the x'-axis, relatively to the stationary x-axis of S, can be represented by oriented rays (solid arrows of Fig.3). Therefore, if the light rays are emitted in the instant of superposition of the origins of S and S', two *light-times* appear: one time corresponds to the forward path of the light, according to the direction of the motion of the source and the other time is in the opposite direction.

So, without introducing LT, we found (Fig.3) a contracted *light-time*

$$t_f = t - \frac{ut}{c} = t\left(1 - \frac{u}{c}\right) = t\left(\frac{c-u}{c}\right), \qquad (10)$$

and a dilated *light-time*

$$t_b = t + \frac{ut}{c} = t\left(1 + \frac{u}{c}\right) = t\left(\frac{c+u}{c}\right). \qquad (11)$$

Therefore, the average *light-time*, in direction of the motion of the source, will be unchanged, that is

$$t = \frac{1}{2}\left[t\left(1 - \frac{u}{c}\right) + t\left(1 + \frac{u}{c}\right)\right]. \qquad (12)$$

Now, at the same manner of a moving light source, also in the longitudinal light clock (Fig.8) we have two opposite light rays along the motion direction of the clock. So, it is necessary to introduce the previous time distortions. Indeed, if we do not introduce the corresponding time distortion, the time $T_{AB}$ required to a light ray to travel the path AB, would be

$$T_{AB} = \frac{L}{c} + \frac{uT_{AB}}{c} = \frac{L}{c-u}, \qquad (13)$$

while for the path BA

$$T_{BA} = \frac{L}{c} - \frac{uT_{BA}}{c} = \frac{L}{c+u}. \qquad (14)$$

But, Eq.(13) and Eq.(14) reveal us an asymmetry of the time which, together to the composition of the velocity of the light with the velocity *u*, is in contrast with the two principles of SR. On the other hand, if we adjust $T_{AB}$ and $T_{BA}$ with the *Lorentz's factor*, contracting or dilating $L$, the asymmetry remains.

Instead, if we introduce the previous time distortions, which appear in Eq.(10) and Eq.(11), we obtain not only $T_{AB} = T_{BA}$, but also the constant velocity $c$. Indeed

$$T_{AB} = \frac{L}{c-u}\left(\frac{c-u}{c}\right) = \frac{L}{c} \quad \text{and} \quad T_{BA} = \frac{L}{c+u}\left(\frac{c+u}{c}\right) = \frac{L}{c} \quad . \tag{15}$$

On the other hand, introducing the *Lorentz's factor*, $T_{AB}$ and $T_{BA}$ dilate in $\gamma T_{AB} = \gamma \frac{L}{c}$ and in $\gamma T_{BA} = \gamma \frac{L}{c}$, requiring necessarily the dilation of $L$ in $\gamma L$.

It is interesting to show that the symmetry of the times is respected also in the *Galilei transformations*. Indeed, in the Galilean scenario, $c+u$ and $c-u$ are respectively the forward and the back velocity of the light that, substituting the velocity $c$ in Eq.(13) and Eq.(14), furnish

$$T_{BA} = T_{BA} = \frac{L}{c} \quad . \tag{16}$$

On the other hand, in lack of the previous time distortions, using Eq.(13) and Eq.(14), we would have

$$T_{AB}' = \frac{L'}{c} + \frac{uT_{AB}'}{c} = \frac{L'}{c-u} \quad \text{and} \quad T_{BA}' = \frac{L'}{c} - \frac{uT_{BA}'}{c} = \frac{L'}{c+u} \quad , \tag{17}$$

where $T_{AB}'$, $T_{BA}'$ and $L'$ are the supposed distorted times, and length, due to the motion of the clock. Hence, the overall time for the light ray, to cover a forward and back path, will be

$$T' = \frac{L'}{c-u} + \frac{L'}{c+u} = \frac{2cL'}{c^2 - u^2} \quad , \tag{18}$$

so $L' = \frac{cT'}{2}\left(1 - \frac{u^2}{c^2}\right)$, but for SR $T' = \frac{T}{\sqrt{1-\frac{u^2}{c^2}}}$, and then we would find a

length contraction, that is

$$L' = L\sqrt{1 - \frac{u^2}{c^2}} \qquad . \qquad (19)$$

## 7. Conclusions

This paper exposes an afterthought, started in previous e-prints [8][9], on the so-called *Lorentz contraction* of the bodies in inertial motion. Essentially the paper is based on this finding: the path of the light, emitted by a moving source and related to a same wave-front, is forward contracted and back dilated, just as happens to its wavelength (Doppler effect). Consequently any other involved length distorts as the path of the light, due the impossibility to verify locally an inertial movement. Therefore, if we consider altogether the path of the light (or its average value) we have a dilation, as it is possible to see from Eq.(3) to Eq.(8). In conclusion, the *Lorentz contraction* is not avoided, but it interests only the front of a moving body, while the body dilates on the whole. In practice, we verify an extension of the Doppler effect which interest also the time and the space.

This Doppler effect is different from the so called *Terrel rotation* [10] which is the distortion that a passing object undergoes at relativistic velocities, when it is observed from a precise point of the space. Instead, our analysis supposes the observer ubiquitous as *Galileo* and *Lorentz transforms* intend. Substantially, the *Terrel rotation* would become a superposition to the time-space Doppler effect.

In this paper, the theme is faced under various aspects, and with more though experiments, to test abundantly the physical coherency of the result and beyond every doubt, that is using the operational method, just as it is used in physics which is not a pure mathematical science.

As it regards the dilated path of muons, which cross the Earth atmosphere, it can be interpreted as dilation of the path which accompanies their relativistic motion, when viewed from the stationary reference system. Consequently, the live-time of muons dilates by the same factor, if measured from the same stationary reference. On the contrary, the supposed contraction of the moving Earth atmosphere, as viewed from muons assumed as stationary reference, cannot be accepted because it is in contrast with the necessary symmetrical and opposite dilation of the moving Earth atmosphere, as viewed from stationary muons.

At least, it must be put again in evidence that the described expansion of the bodies is not symmetrical about their density, for which an homogeneous body that travels at relativistic velocity would have a more massive frontal part and an evanescent tail, along the direction of its motion. Under this aspect  relativistic particles, spherical and isotropic, would appear as in Fig.(9)  as it regards the density.

**Figure captions**

**Fig. 1** Schematic drawing of a transversal light clock, in uniform translation motion with relativistic velocity *u* (clock 2), relatively to an equal clock considered stationary (clock 1).

**Fig. 2** Wave-fronts of the light emitted isotropically from a point source, when viewed locally in the empty space. The solid arrows, representing the light rays along a Cartesian axis, measure the time interval elapsed from the initial instant of emission.

**Fig. 3** Wave-fronts of the light emitted from a point source moving at relativistic velocity *u* in the empty space. The solid arrows represent the light rays, forward and back along the direction of motion, and measuring the times elapsed from the instant of superposition of the source with the origin of the stationary Cartesian reference.

**Fig. 4** Wave-fronts of the light emitted isotropically from a point source, integral and concentric with a transparent rigid hollow sphere, when viewed locally in the empty space.

**Fig. 5** Impossibility of the previous rigid shell to move symmetrically contracted (item 1), unchanged (item 2) or symmetrically dilated (item 3) in the empty space, relatively to the stationary reference S. The intersections among the shell and the wave-fronts of the light cannot be viewed locally.

**Fig. 6** Impossibility of contraction of the moving rigid shell of Fig.3, in concert with the contraction of the space, for the impossibility of the wave fronts of the light to be dragged by the motion of the source.

**Fig. 7** Relativistic dilation of the moving shell of Fig.3, in concert with the dilation of the space, avoiding the intersections of the wave-fronts of the light with the shell are .

**Fig. 8** Schematic drawing of a longitudinal light clock in uniform translation motion at relativistic velocity *u* (clock 2), compared to an equal clock assumed stationary (clock 1).

**Fig.9** Artistic view of the asymmetrical distortion of the density of a spherical particle, of uniform mass when it is at rest, moving at relativistic velocity *u*.

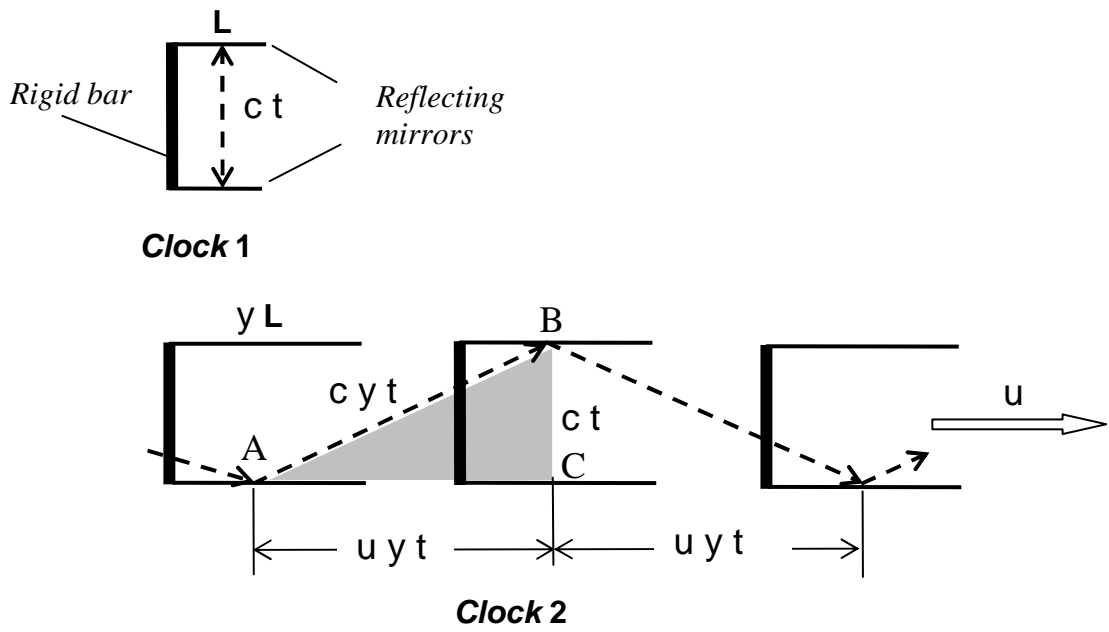

**FIG. 1**

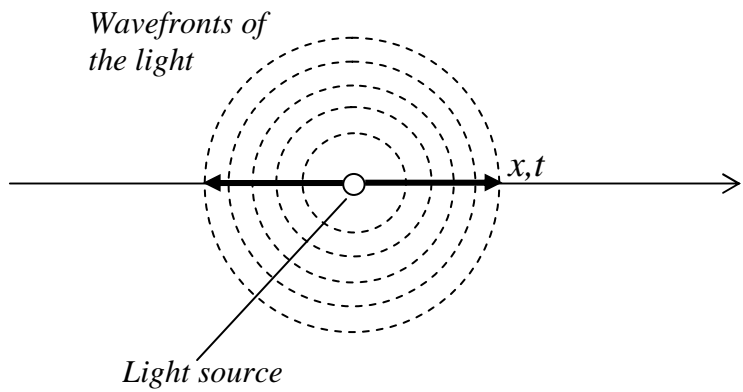

**FIG. 2**

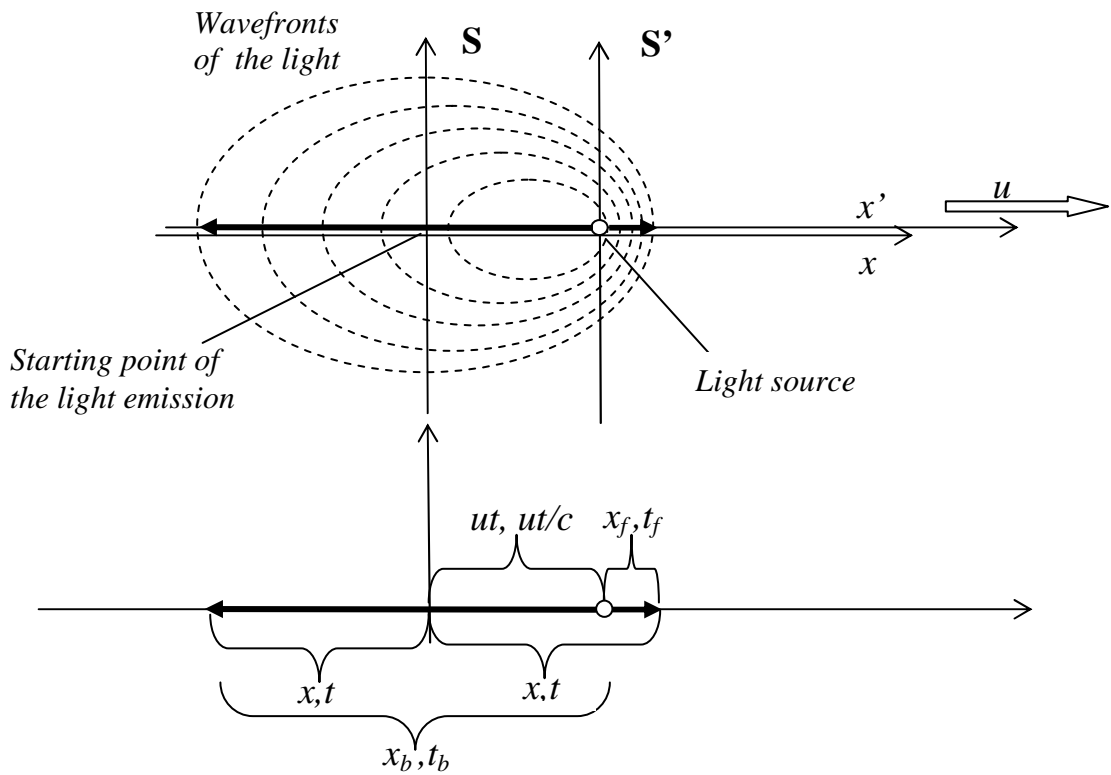

**FIG. 3**

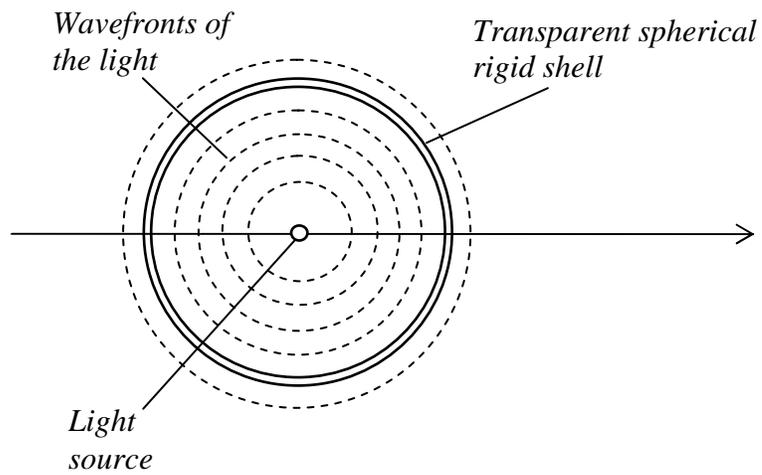

**FIG. 4**

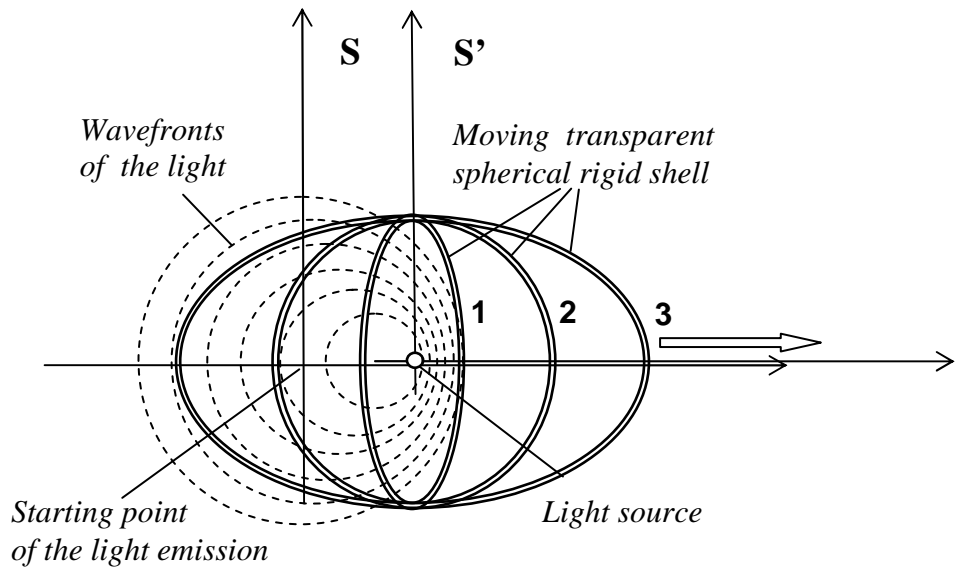

**FIG. 5**

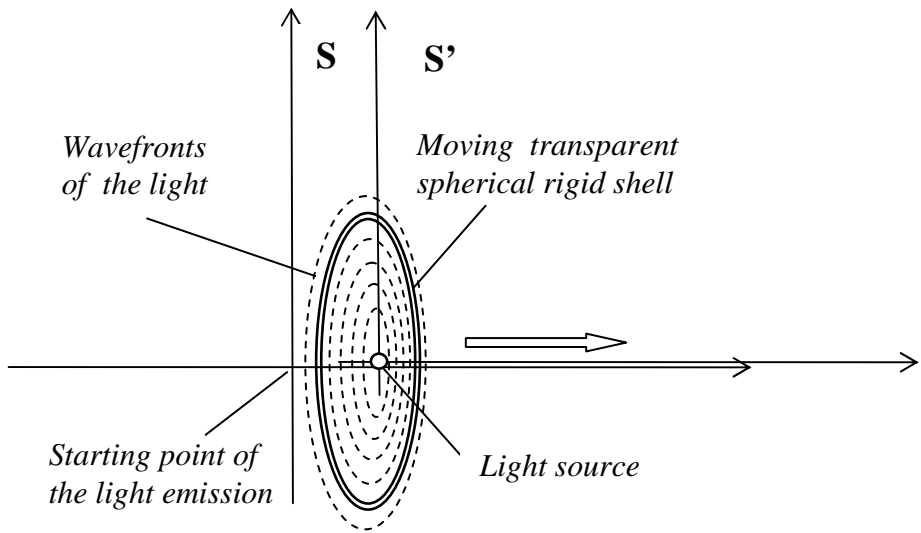

**FIG. 6**

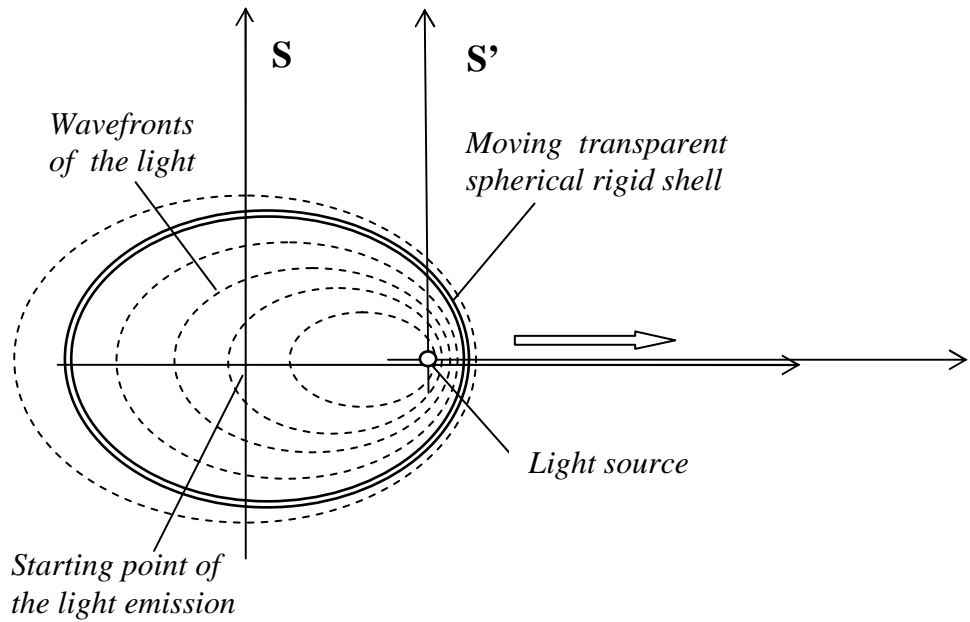

**FIG. 7**

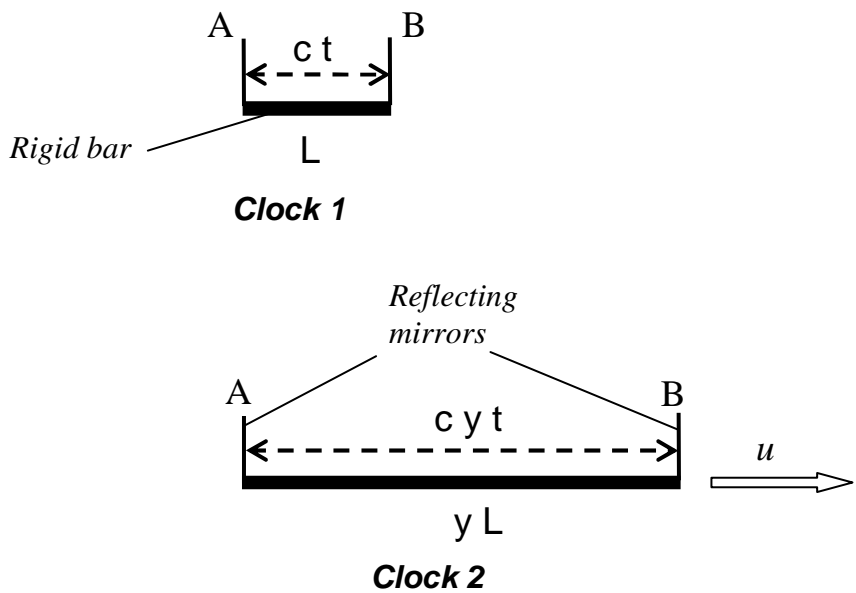

**FIG. 8**

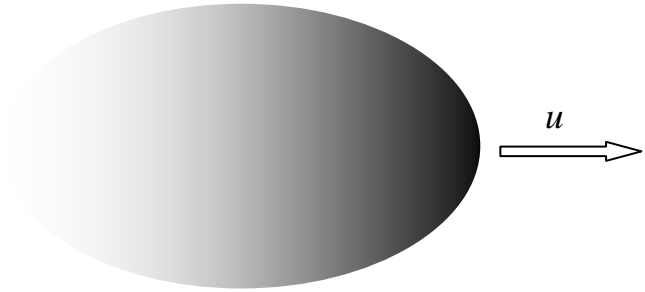

**FIG. 9**